# Superconductivity at 25K under hydrostatic pressure for FeTe$_{0.5}$Se$_{0.5}$ superconductor

Rajveer Jha[1], Rayees A. Zargar[2], A. K. Hafiz[2], H. Kishan[1] and V.P.S. Awana[1,*]

[1]CSIR-National Physical Laboratory, Dr. K.S. Krishnan Marg, New Delhi-110012, India

[2]Department of Physics, Jamia Millia Islamia, New-Delhi, 110025

We report the impact of hydrostatic pressure on the superconductivity and normal state resistivity of FeTe$_{0.5}$Se$_{0.5}$ superconductor. At the ambient pressure the FeTe$_{0.5}$Se$_{0.5}$ compound shows the superconducting transition temperature T$_c^{onset}$ at above 13K and T$_c^{\rho=0}$ at 11.5K. We measure pressure dependent resistivity from 250K to 5K, which shows that the normal state resistivity increases initially for the applied pressures of up to 0.55GPa and then the same is decreased monotonically with increasing pressure of up to 1.97GPa. On the other hand the superconducting transition temperatures (T$_c^{onset}$ and T$_c^{\rho=0}$) increase monotonically with increasing pressure. Namely the T$_c^{onset}$ increases from 13K to 25K and T$_c^{\rho=0}$ from 11.5K to 20K for the pressures range of 0-1.97GPa. Our results suggest that superconductivity in this class of Fe-based compounds is very sensitive to pressure as the estimated pressure coefficient dT$_c$(onset)/dP is ~ 5.8K/GPa. It may be suggested that FeTe$_{0.5}$Se$_{0.5}$ superconductor is a strong electron correlated system. The enhancement of T$_c$ with applying pressure is mainly attributed to an increase of charge carriers at Fermi surface.

**Introduction**

The discovery of Fe-based superconductor LaFeAsO$_{1-x}$F$_x$ with superconducting transition temperature (T$_c$) above 26K has attracted much attention of condensed matter scientific community [1]. In this direction, other iron- Pnictides and Fe-chalcogenide superconductors have been developed such as LnFeAsO$_{1-x}$F$_x$ (Ln=Sm, Nd, Pr, and Ce 1111-family) [2-5] with T$_c$ of up to 55K, T$_c$=38K in Ba$_{1-x}$K$_x$Fe$_2$As$_2$, [6], T$_c$=18K in LiFeAs, [7] and T$_c$=8.0K in Fe$_{1+x}$Se [8]. Interestingly all the Fe-based superconductors have layered structure such as FeAs layer in pnictides and FeSe layer in chalcogenide, as similar to that of CuO$_2$ layer in high temperature cuprate superconductors [9]. The parent compounds of all Fe-based superconductors show the spin-density-wave (SDW) transition accompanied with a structural change from tetragonal to monoclinic at low temperatures [1-6]. The superconductivity can be induced with elemental doping in the parent compounds [1-11]. In FeSe system, the superconductivity can be induced up to 8K with either excess of Iron or the deficiency of Selenium [12, 13]. Under hydrostatic pressure the superconductivity of FeSe superconductor has been increased up to 27K [14]. The $FeTe_{1-x}Se_x$ chalcogenide system has the simplest crystal structure among the other iron-based superconductors that gives the positive discrimination to its experimental and theoretical studies related to on site chemical substitution and the impact of high pressures [15, 16]. The impact of high pressure on $FeSe_{1-x}Te_x$ compounds showed that the crystal structure change from tetragonal (P4/nmm) phase to the orthorhombic (Pbnm) at high pressures and low temperatures [17-19]. Also the pressure dependent superconducting properties have been studied for the $FeTe_{0.5}Se_{0.5}$ system, which showed the improvement in $T_c$ from 14K to 22K for the pressure of 25.3kbar [20].

The pressure effect is one of the powerful tools to study the superconducting properties by effectively changing the primary parameters such as the lattice constants through bond lengths and angles without doping, which may improve the electronic density of states at the Fermi surface. Here we report the pressure dependent resistivity from 250K down to 5K for the various applied pressures from 0 to 1.97GPa for the $FeTe_{0.5}Se_{0.5}$ superconductor. The $FeTe_{0.5}Se_{0.5}$ compound has been synthesized by solid state reaction rout via vacuum encapsulation, which is crystallized in tetragonal structure with space group P4/nmm. The normal state resistivity decreases with increasing pressure and the superconducting transition temperature $T_c$ is enhanced up to 25K at 1.97GPa pressure. Our results are qualitatively in confirmation with ref. 20.

**Experimental:**

The bulk polycrystalline $FeTe_{0.5}Se_{0.5}$ Sample was synthesized through standard solid state reaction route via vacuum encapsulation. The highly purity chemicals Fe, Se, and Te are weighed in the stoichiometric ratio and ground thoroughly Glove box having pure Argon atmosphere. The mixed powder was subsequently pelletized and then encapsulated in an evacuated ($10^{-3}$ Torr) quartz tube. The encapsulated tube is then heated at 750 $^o$C for 12 hours and slowly cooled to room temperature. The X-ray diffraction (XRD) was taken at room temperature in the scattering angular (*2θ*) range of 10$^o$-80$^o$ in equal *2θ* step of 0.02$^o$ using *Rigaku Diffractometer* with *Cu K$_α$* (*λ* = 1.54Å). Rietveld analysis was performed using the standard *FullProf* program. The pressure dependent resistivity measurements were performed by using HPC-33 Piston type pressure cell with Quantum design DC resistivity option on Physical Property Measurements System (*PPMS*-14T, *Quantum Design*). Hydrostatic pressures were generated on the sample by a BeCu/NiCrAl clamped piston-cylinder cell, which was immersed in a fluid pressure transmitting medium of Fluorinert (FC70:FC77=1:1) in a Teflon cell. Four probe resistivity methods is used to measure resistivity of the sample.

**Results and discussion:**

The room temperature observed and Reitveld refined XRD pattern of $FeTe_{0.5}Se_{0.5}$ compound are shown in Fig. 1. The compound is fitted in tetragonal structure with space group

P4/*nmm*. Within the XRD detection limit no impurity phase is seen in the main phase of the compound. Reitveld refined lattice parameter are found to be a = 3.79(1)Å and c = 6.02 (2)Å. Inset of the Fig.1 shows the unit cell of the FeTe$_{0.5}$Se$_{0.5}$ compound, which has been drawn from the refined parameters. It can be seen that the Fe is located at coordinate positions (0.75, 0.25, 0) and Se/Te at (0.25, 0.25, 0.272). Our results are quite similar to the earlier reported results [15, 16].

Fig.2 represents the temperature dependent resistivity ρ(T) for the FeTe$_{0.5}$Se$_{0.5}$ compound from 250K down to 5K at the various applied pressure from 0 to 1.97GPa. It can be seen from Fig.2 that the normal state resistivity (ρ$^{250K}$) decreases slightly with increasing pressure of up to 0.55GPa. The normal conduction ρ(T) is of slightly metallic nature and the same improves with pressure. The decrease in normal state resistivity with improved metallic behavior may take place due to the alignment of distorted structure of this system resulting in increase of number of charge carriers at the Fermi level [17]. Figure 3 shows the enlarged part of the ρ(T) under pressure from 30K to 10K to see the variation of superconducting transition temperature with applied pressure. At zero pressure T$_c^{\rho=0}$ is at 11.5K, which increases to 12.5K at 0.35GPa pressure. Interestingly, with small increase in pressure from 0.35GPa to 0.55GPa, the T$_c^{\rho=0}$ is increased from 12.5K to 16K. For further higher pressures the T$_c^{\rho=0}$ increases gradually to 20K with increasing pressure to 1.97GPa. It is also seen that with increasing applied pressure the superconducting transition width (T$_c^{onset}$ - T$_c^{\rho=0}$) increases and is maximum (~5K) for the highest possible applied pressure of 1.97GPa.

Figure 4 shows the temperature derivative of resistivity (dρ/dT) for the FeTe$_{0.5}$Se$_{0.5}$ compound at the various applied pressure from 0 to 1.97GPa. The midpoint of the superconducting transition T$_c$ can be seen clearly from the dρ/dT peak. Height of the transition peak is highest for the applied pressure 0.55GPa, which decreases for the further higher applied pressures. It can be clearly seen that with increasing pressure the width of the dρ/dT peak increases. This shows that broadening of superconducting transition takes place with pressure. Also secondary peaks are seen for higher pressures of 1.68GPa and 1.97GPa. The sufficient increase in broadening of transition with secondary peaks at higher pressures is indicative of phase separation. Figure 5 shows the summarized results for the three different region of superconducting Transition temperature T$_c^{\rho=0}$, T$_c^{mid}$, and T$_c^{onset}$ vs Pressure for the FeTe$_{0.5}$Se$_{0.5}$ compound. It can be seen from form Fig. 5 that the superconducting transition temperature (T$_c^{\rho=0}$) increases gradually from 11.5K- to 20K for the applied pressure of 0-1.97GPa. The obtained highest T$_c^{onset}$ is 25K for the applied pressure of 1.97GPa. The pressure coefficient d T$_c^{onset}$/dP of FeTe$_{0.5}$Se$_{0.5}$ is estimated to be ~5.8K/GPa. This value is close to earlier reported values of pressure coefficient for the iron-based chalcogenide superconductors [14, 20]. Increase in the T$_c$ with applying pressure for FeTe$_{0.5}$Se$_{0.5}$ system is due to the increase in number of density of state at the Fermi surface and enhanced electron–phonon coupling constant outweighing the lattice stiffening effect. It would be possible to attain a higher T$_c$ by applying still higher pressures. In our study, at present we are limited to a maximum of 2GPa pressure.

In earlier studies on iron based compounds, it has been suggested that the superconductivity is correlated with the suppression of SDW ordering, which can be achieved by doping such as in $PrFeAsO_{1-x}F_x$, $NdFeAsO_{1-x}F_x$ and $Ba_{1-x}K_xFe_2As_2$ or by applying external pressure as well [2, 3, 6]. In the binary FeSe superconductor the largest positive pressure coefficient of $T_c$ has been discussed in terms of spin instabilities [14]. On the other hand, the large pressure coefficient of $FeTe_{0.5}Se_{0.5}$ compound is related to the large atomic size of Te, which seems sensible to the crystal lattice vibrations [20]. With applying pressure decrease in normal state resistivity suggests that the structure gets aligned and the grain boundary contributions are reduced. It may also be suggested that $FeTe_{0.5}Se_{0.5}$ superconductor is a strong electron correlated system invoking increased hybridization of Fe and Se under pressure. .In our case the normal state resistivity is suppressed with the applied pressure accompanied with improved metallic behavior.

**Conclusion:**

In conclusion, we studied the impact of hydrostatic pressure on superconductivity of $FeTe_{0.5}Se_{0.5}$ superconductor. The observed pressure coefficient $dT_c^{onset}/dP$ is 5.8K/GPa, which is comparable to earlier reported values for the similar superconducting compounds. The much higher pressure coefficient may happen due to increase of density of states at Fermi level with applied pressure.

**Acknowledgement:**

Authors would like to thank their Director NPL India for his keen interest in the present work. This work is financially supported by *DAE-SRC* outstanding investigator award scheme on search for new superconductors. Rajveer Jha acknowledges the *CSIR* for the senior research fellowship. H. Kishan thanks *CSIR* for providing Emeritus Scientist Fellowship.

**Reference:**

1. Y. Kamihara, T. Watanabe, M. Hirano, and H. Hosono, J. Am. Chem. Soc. **130,** 3296 (2008).
2. X. H. Chen, T. Wu, G. Wu, R. H. Liu, H. Chen, and D. F. Fang: Nature **453,** 761 (2008).
3. G. F. Chen, Z. Li, D. Wu, G. Li, W. Z. Hu, J. Dong, P. Zheng, J. L. Luo, and N. L. Wang, Phys. Rev. Lett. **100**, 247002 (2008).
4. Z. A. Ren, J. Yang, W. Lu, W. Yi, X.-L. Xhen, Z.-C. Li, G.-C. Che, X.-L. Dong, L.-L. Sun, F. Zhou, and Z.-X. Zhao, Europhys. Lett. **82**, 57002 (2008).
5. Z.A. Ren, J. Yang, W. Lu, W. Yi, G.-C. Che, X.L. Dong, L.L. Sun, Z.X. Zhao, Materials Research Innovations **12**, 105 (2008).
6. M. Rotter, M. Tegel, and D. Johrendt, Phys. Rev. Lett. **101**, 107006 (2008).
7. X. C. Wang, Q. Q. Liu, Y. X. Lv, W. B. Gao, L. X. Yang, R.C. Yu, F. Y. Li, and C. Q. Jin, Solid State Commun. **148,** 538 (2008).

8. F. C. Hsu, J. Y. Luo, K. W. Yeh, T. K. Chen, T. W. Huang, P. M. Wu, Y. C. Lee, Y. L. Huang, Y. Y. Chu, D. C. Yang, and M. K. Wu, Proc. Natl. Acad. Sci. U.S.A. **105,** 14262 (2008).
9. J. G. Bednorz and K. A. Muller, Z. Phys. B **64**, 189 (1986).
10. C. de la Cruz, Q. Huang, J. W. Lynn, J. Li, W. Ratcliff II, J. L. Zarestky, H. A. Mook, G. F. Chen, J. L. Luo, N. L. Wang & P. Dai, Nature **453**, 899 (2008).
11. D. J. Singh and M.-H. Du, Phys. Rev. Lett. **100,** 237003 (2008).
12. J. Dong, H. J. Zhang, G. Xu, Z. Li, G. Li, W. Z. Hu, D. Wu, G. F.Chen, X. Dai, J. L. Lou, Z. Fang, and N. L. Wang: Europhys. Lett. **83**, 27006 (2008).
13. T. M. McQueen, Q. Huang, V. Ksenofontov, C. Felser, Q. Xu, and R. J. Cava, Phys. Rev. B **79**, 014522 (2009).
14. Y. Mizuguchi, F. Tomioka, S. Tsuda, and T. Yamaguchi, Appl. Phys. Lett. **93**, 152505 (2008).
15. M. H. Fang, H. M. Pham, B. Qian, T. J. Liu, E. K. Vehstedt, Y. Liu, L. Spinu, and Z. Q. Mao: Phys. Rev. B **78**, 224503 (2008).
16. V. P. S. Awana, A. Pal, A. Vajpayee, M. Mudgel, H. Kishan, M. Husain, R. Zeng, S Yu, Y. F. Guo, Y. G. Shi, K. Yamaura and E. T. Muromachi, J. Appl. Phys. **107**, 09E128 (2010).
17. R. S. Kumar, Y. Zhang, V.P.S. Awana, A. Kumar, Y. Xiao, S. Sinegeikin, P. Chow, A. Cornleius, C. Chen and Y. Zhao, to appear in J. Phys. Chem. Solids (2014).
18. R. S. Kumar, Y. Zhang, S. Sinogeikin, Y. Xiao, S. Kumar, P. Chow, A. L. Cornelius, and C. Chen, J. Phys. B **113**, 12597 (2010).
19. P. S. Malavi, S. Karmakar, N. N. Patel, H. Bhatt, and S. M. Sharma, J. Phys. Condens, Matter, **26,** 125701 (2014).
20. C. L. Huang, C. C. Chou, K. F. Tseng, Y. L. Huang, F. C. Hsu, K. W. Yeh, M. K. Wu, and H. D. Yang, J. Phys. Soc. Japan,**78**, 084710 (2009).

**Figure Captions**

**Figure 1:** Reitveld fitted XRD pattern at room temperature Observed (*open circles*) and calculated (*solid lines*) of FeTe$_{0.5}$Se$_{0.5}$ compound.

**Figure 2:** Temperature dependent Resistivity (ρ Vs T) at various applied pressures in the temperature range 250K-5K for FeTe$_{0.5}$Se$_{0.5}$ compound.

**Figure 3:** Resistivity versus temperature (ρ Vs T) plots for FeTe$_{0.5}$Se$_{0.5}$ compound, at various applied pressures in the temperature range 30K-10K.

**Figure 4:** Temperature derivative of Resistivity for FeTe$_{0.5}$Se$_{0.5}$ compound, at various applied pressures in the temperature range 30K-10K.

**Figure 5:** $T_c^{\rho=0}$, $T_c^{mid}$, and $T_c^{onset}$ versus pressure plots for FeTe$_{0.5}$Se$_{0.5}$ compound.

Figure 1

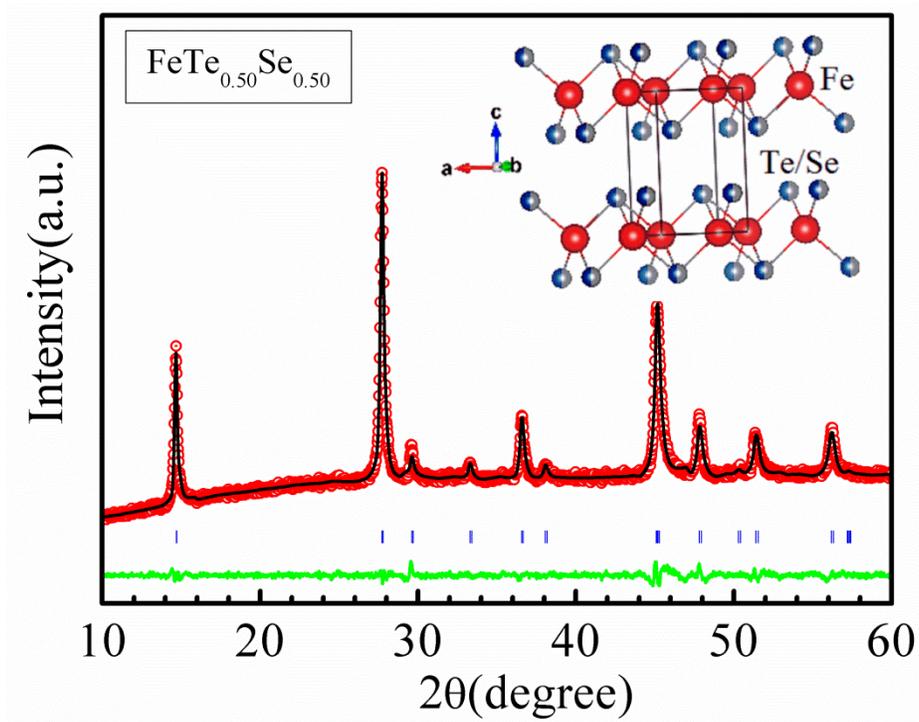

Figure 2

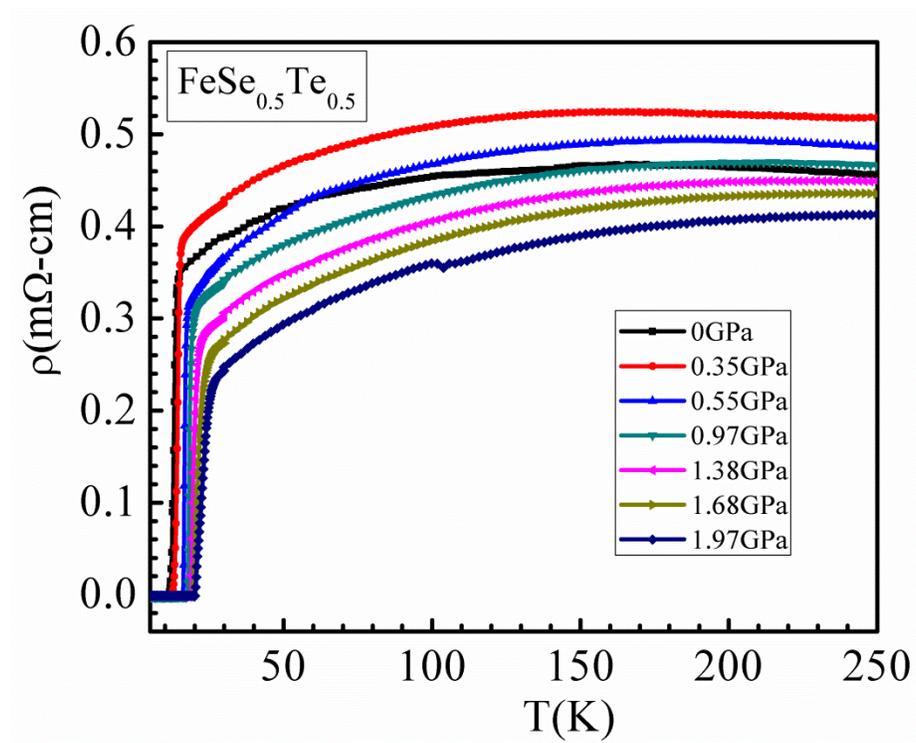

Figure 3

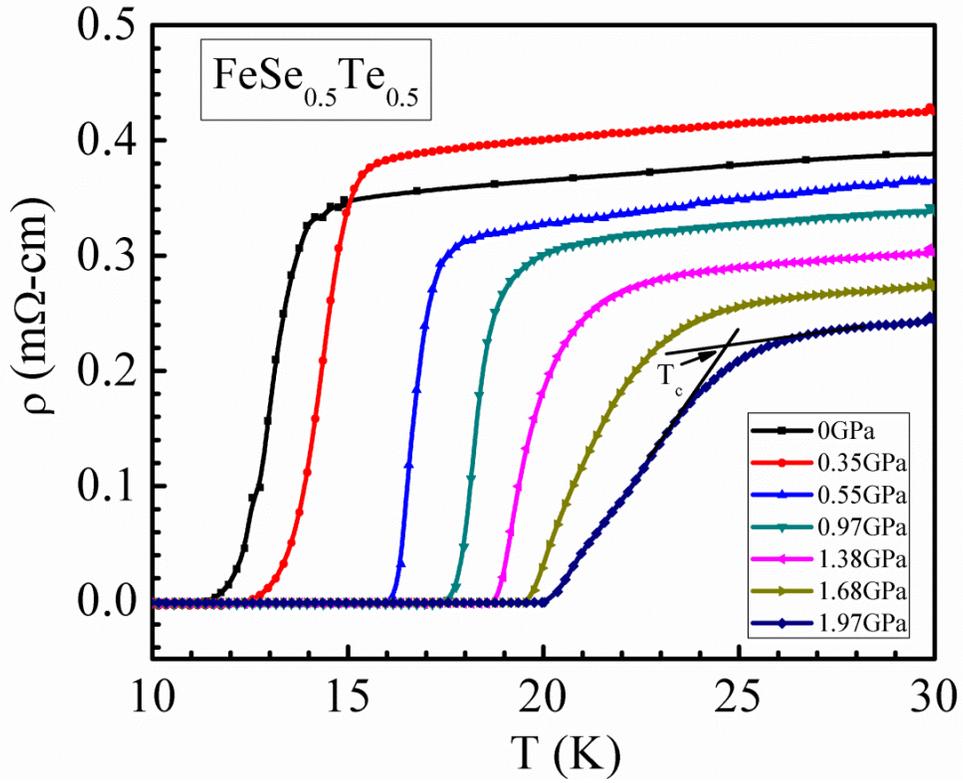

Figure 4

Figure 5

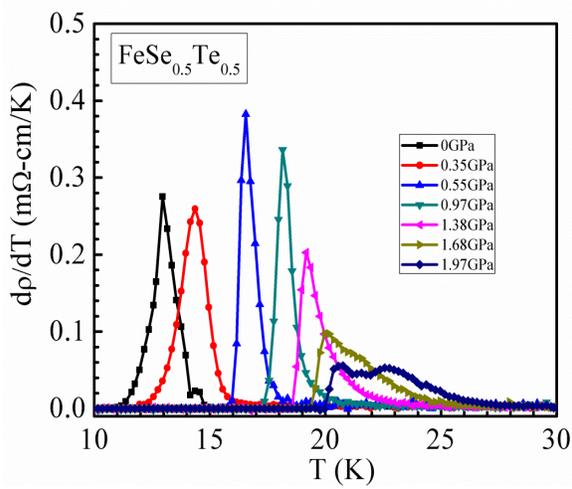
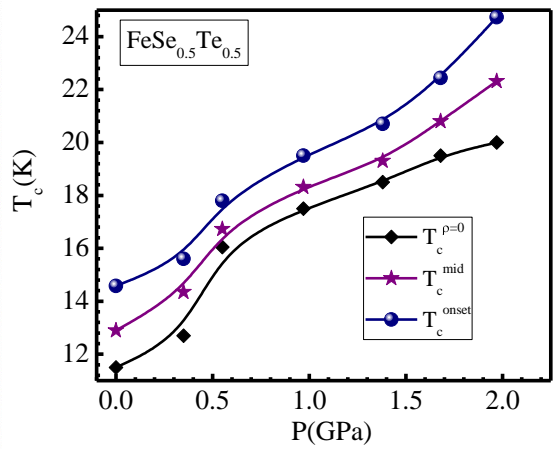